**Title**: A simulation model for longitudinal HIV viral load and mother-to-child-transmission in pregnant and postpartum women

**Authors:** Maia Lesosky[1], Janet Raboud[2], Tracy Glass[1], Elaine Abrams[3], Landon Myer[1]

**Affiliations:**
1. Division of Epidemiology & Biostatistics, School of Public Health & Family Medicine, University of Cape Town
2. University Health Network, Toronto, Canada
3. ICAP, Mailman School of Public Health, Columbia University, New York, USA; College of Physicians & Surgeons, Columbia University, New York, USA

**Running head:** Simulation model of HIV viral load in pregnancy

**Funding:** Eunice Kennedy Shriver National Institute of Child Health & Human Development of the National Institutes of Health under award number R21HD093463. The content is solely the responsibility of the authors and does not necessarily represent the official views of the National Institutes of Health.

**Abstract:**

This manuscript describes the model specification, including input and output measures, dependencies, and structure for VLSiM-PPW, a Monte Carlo type model for HIV adherence, viral load and vertical transmission in pregnant and breastfeeding women.

# 1 INTRODUCTION

Viral load monitoring among people living with HIV (PLWH) is recognised as the gold standard for monitoring of treatment adherence and monitoring the development of treatment resistance [1]. There has been a rapid and wide scale up of viral load monitoring in low- and middle- income countries globally [2]. There are global concerns about adherence to ART in pregnant and breastfeeding women and subsequent elevated viral load (eVL) and mother-to-child-transmission (MTCT) risk [3]. Intensified viral load monitoring for pregnant and breastfeeding women has been proposed in some guideline recommendations but not evaluated systematically [4-6].

A simulation model for longitudinal HIV viral load (VL) during pregnancy and breastfeeding (VLSiM-PPW) has been developed to help evaluate guidelines based VL monitoring across a variety of individual and health system parameters [7]. The motivation and need for research in this area is considerable [3,8] and this model was created in response.

This manuscript details the model specification, including input and output measures, dependencies, and outlines the calibration and validation approach.

# 2 METHODS

**Overview of model specification**

Figure 1 describes the high-level model structure. A number of tables and schematics are presented to assist with understanding the model structure, as well as following text detailing the model structure and assumptions.

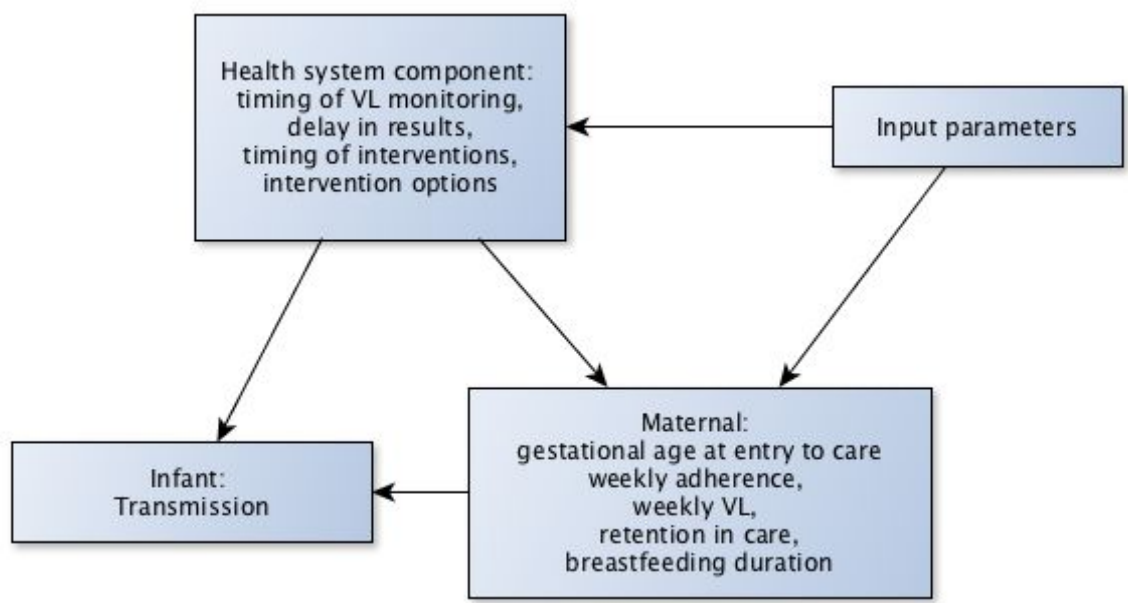

**Figure 1:** Overview of model components with key constructs modelled

This is an individual, Monte Carlo type simulation model [9] simulating longitudinal antiretroviral therapy (ART) adherence, retention in care, HIV VL, breastfeeding and vertical transmission risk from conception until two years postpartum on a weekly time step. In addition to a number of individual parameters and outcomes, parameters at the level of health services are also modelled, including the possibility to: monitor VL, apply interventions of varying effectiveness and model the delay between timing of VL monitoring and return of test results. The overall objective of the model is to adequately simulate ART adherence, VL, and MTCT risk in pregnant and breastfeeding women in sub-Saharan Africa (SSA), in order to inform optimal VL monitoring schedules for early detection of raised VL during this period. The main outcome measures of interest are: the proportion of women becoming viremic over time, the proportion of women eligible for VL monitoring, the number of monitoring VL measurements taken, the median (IQR) of viral load at time of measurement, the cumulative viremia (cVL) experienced before detection, the number of interventions undertaken and the total number of vertical transmissions.

Individuals can enter the simulation continuing, or initiating, ART and can be subject to different probabilities of adherence, different monitoring schedules and different impacts of intervention. Women can enter care at different gestational ages, and have a weekly risk of being lost to care. Detailed description of each of the model components can be found below.

**Description of model components**

**Temporal phases** As this model is oriented towards pregnant and postpartum women (PPW) the model is compartmentalized into four natural periods reflecting stages of pregnancy and postpartum: Period 1: from conception until entry into antenatal care (ANC), Period 2: from ANC until delivery, Period 3: from delivery until the cessation of breastfeeding (BF), Period 4: from the cessation of BF until the end of 24 months postpartum (Figure 2). Individuals are modelled through all four periods in a contiguous manner and on a consistent weekly time step. Although the model is focused on the use and utility of VL monitoring during antenatal and early infant care, Period 1 is necessary to completely model vertical transmission, as infants are at risk of HIV acquisition from early post conception through to the end of breastfeeding. Certain considerations and interventions are only available in specific periods. For example, infant prophylaxis and interventions aimed at breastfeeding duration are only available in Period 3. Most importantly, VL monitoring and any intervention is only possible during Periods 2 and 3, where women are assumed to be in antenatal or mother-child based services. As the model incorporates the possibility of lost to follow up (or non-engagement in care) (LTFU), it is possible for women/infants to be modelled in Periods 2 and 3 without possibility of VL monitoring.

| Simulation time periods | Conception until entry to ANC | ANC entry until delivery | Delivery until end of BF | End BF until 2y PP (whichever first) |
|---|---|---|---|---|
| VL monitoring possible | | VL monitoring while in care | | |
| Maternal interventions applied | | Maternal interventions while in care | | |
| Maternal/infant LTFU possible | | Weekly LTFU risk | | |
| Infant interventions applied | | | Infant interventions | |
| MTCT calculated | Weekly MTCT transmission risk | | | |

**Figure 2:** Temporal phasing of simulation model.

**Model inputs**

Model input parameters were considered in three classes:
1. Fixed: these are internal and/or biological variables that are either fixed by definition (i.e., floor or ceiling thresholds for computational stability) or will be fixed after internal validation and will not change for sensitivity, calibration or data analyses. These are mentioned throughout, but primarily discussed in the validation & calibration sections.
2. Optimized: these are internal variables that have been through the calibration process and are subject to goodness-of-fit evaluation across a variety of data sources.
3. Input: these are parameters, usually population based, that will be selected and/or varied depending on the context of the question or population being modelled. These are the model inputs relevant to most investigators and will be the model inputs used to parameterize the model for evaluation of specific scientific questions. These input parameters are described in Table 1.

**Maternal adherence** We use the term "adherence" or "maternal adherence" to refer to the spectrum of retention and adherence patterns that may be present after a woman has been diagnosed and initiated to ART. As is well described empirically [10-12], adherence to ART, or to any chronic medication, is a complex behavioral phenomenon. Individuals may choose to consistently take medication or not for a variety of reasons, for example, perceived stigma or side-effects, and they may accidentally fail to take medication at different times for an unrelated set of causes, for example, a lost or forgotten bag containing medication. Both individual behavioral characteristics and stochastic events can impact the week-to-week levels of ART circulating in an individual's bloodstream, and subsequently the levels of circulating virus. Broadly speaking these patterns of behavior and access (consistency of pill taking), in the absence of developing HIV drug resistance, lead to either well controlled VL or to poorly controlled VL (note that HIV drug resistance is not currently modelled directly but may be reflected by varying other parameters).

This simulation model does not mimic the complex multifactorial causes of variation in adherence, but models the observed variability in VL under a minimal set of key parameters. This is done by assuming and assigning each individual a "baseline adherence" category, which is calibrated to observed rates of viremia during pregnancy and postpartum, but can be altered by changing input settings. This initial category generates a baseline probability using a random draw from a Gaussian mixture distribution. The baseline adherence probability is allowed to change over time (eg different baseline adherence during pregnancy compared to postpartum) as well as undergo an attenuation effect on either side of delivery. The attenuation effect and different pre- and post-delivery baseline adherence probabilities are controlled by simple binary switches. The updated "weekly adherence" probability is predicted from a linear model incorporating baseline adherence probability, any intervention effect currently in action, the previous weeks VL and how long the individual has been adherent or not, as well as a stochastic additive noise effect. This resulting weekly adherence probability is categorised as below or above the threshold for 'adherence', and then informs the estimation of the current VL value.

**Table 1:** Model inputs altering simulation population characteristics and details of simulation setup

| Input parameter | Description | Range |
|---|---|---|
| **Characteristics of population of pregnant and BF women entering care** | | |
| Mean gestational age (weeks) at entry to ANC | Gestational age simulated from a mixture distribution using the input mean GA as primary peak mean. Rejection sampling used to ensure GA does not fall out of range boundary (8 weeks - delivery) and is prior to delivery. | 8 – 38 weeks (or delivery) |
| Mean duration of BF (weeks) | BF duration for individuals is simulated from a mixture distribution taking input mean as primary peak. Rejection sampling used to ensure BF duration is non-negative. | 0 – 96 weeks |
| Percent of women initiating BF | Taken as input | 0 – 100% |
| Percent of women continuing ART | Taken as input | 0 – 100% |
| Total percent of women LTFU antenatally | Converted into a weekly, individual probability of LTFU. Every week, for each individual a Bernoulli random variable is generated with weekly LTFU probability, if a success then women is considered LTFU from the following week. | 0 – 100% |
| Total percent of women LTFU postnatally | Converted into a weekly, individual probability of LTFU. Every week, for each individual a Bernoulli random variable is generated with weekly LTFU probability, if a success then women is considered LTFU from the following week. | 0 – 100% |
| Percent of women with very poor/no ART adherence | Can be set above zero to fix a percentage of simulated women who never progress towards or maintain viral suppression. | 0 - 100% |
| Percent of women with full adherence | Can be used to set simulations with some percentage of women who maintain viral suppression throughout. | 0 – 100% |
| Percent of women with poor/mixed ART adherence | The remaining proportion of women who are neither classed as 'full' or 'never' adherent. Sum of the three must be 100%. | 0 – 100% |
| **Characteristics of health system** | | |
| VL monitoring delay | Average time in weeks from when a VL is measured to when the result is available to be fed back to the woman | 0 - 16 weeks |
| VL monitoring strategy | Taken as input (see Table 2 and Supplement Table 1) | Varies |
| Maternal intervention strategy | Can set duration of effect and percent of women it will be effective in. | Varies |
| Infant intervention strategy | Three fixed options: none, 'regular' and 'enhanced'. See text for details. | Varies |

BF: breastfeeding; GA: gestational age; LTFU: lost to follow up; ART: antiretroviral therapy; ANC: antenatal care

There are three categories of baseline adherence: "complete adherence", "mixed adherence" and "non-adherence", the proportion of women in each category is an input parameter. Complete adherence will result in rapid viral suppression and sustained maintenance of viral suppression, with low probability of transient or long term viremia. Similarly, non-adherence will result in a lack of viral suppression (sustained viremia) with a low probability of meeting low VL targets. The mixed adherence group represents a group that may experience periodic episodes of non-adherence, or may simply have adherence probabilities that fluctuate with a larger magnitude, hence causing VL to fluctuate as well. Maximum VL for each individual is established via a viral set point simulated as "pre-ART VL" in the simulation initiation. Figure 3 provides a schematic of modeled adherence and factors influencing adherence.

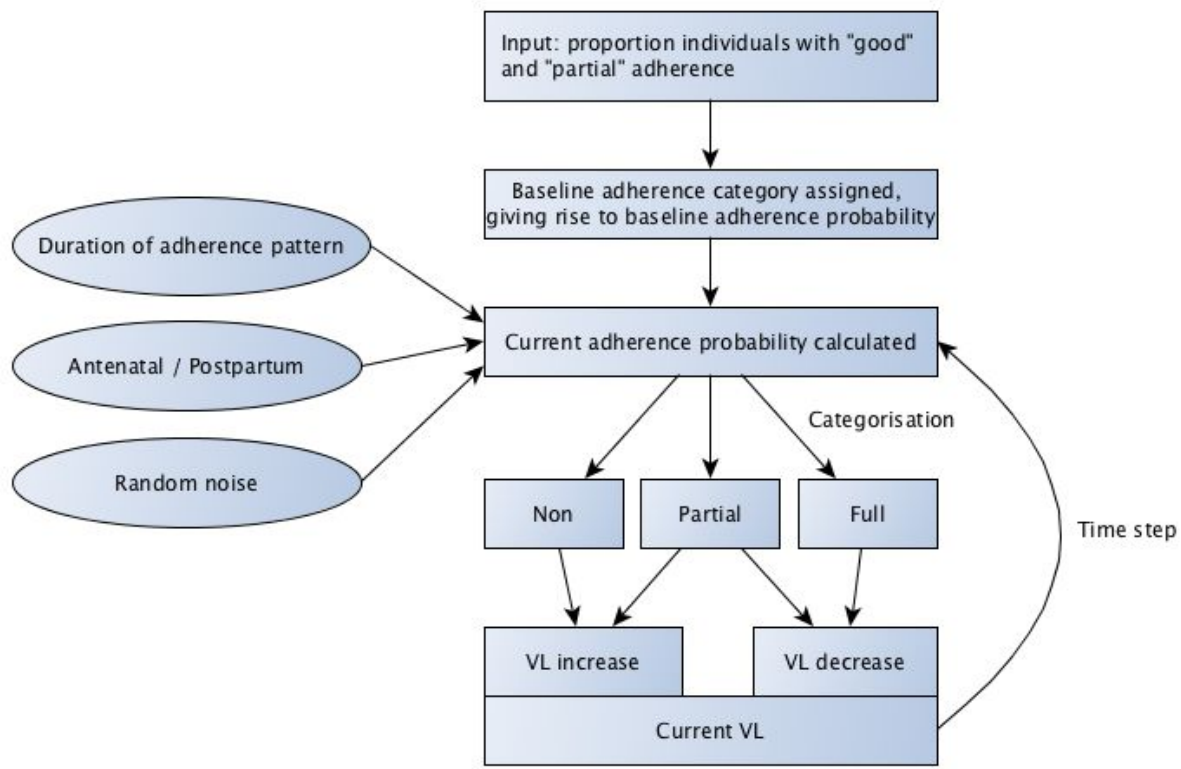

**Figure 3:** Schematic representation of modeled adherence and factors influencing adherence

**VL trajectories** VL values are initiated at the point of conception, after the assignment of continuing or initiating ART status and initialization of baseline adherence category. The initialization VL represents an individual's mean expected VL under no change in adherence probabilities or biology. The distributions of initial VL have been based on empirical data, and are subject to sensitivity analyses. There are three possible models for simulation of individual weekly VL, characterized by their mean trend: stationary, decreasing and increasing.

A stationary distribution for VL is used prior to ART initiation and/or when individuals reach the VL set point or viral suppression at less than or equal to 50 copies/mL. The pattern of VL change over time in the stationary distribution is one of random noise around a zero-slope trajectory. Random noise is magnitude dependent above 1000 copies/mL, at $\gamma* \log10(VL)$ [13,14] where $\gamma$ is a modifiable internal parameter subject to calibration.

A decreasing trajectory, applied post ART initiation, and if/when individuals return to ART adherence after a period of non-adherence, demonstrates a mean decreasing trend of VL over time, including random noise. An increasing trajectory is used when individuals are non-adherent or lost to follow-up and no longer in care. The duration spent in each of these trajectories depends on the weekly adherence probability. As with the stationary distribution, the random noise added to the current VL calculation is magnitude dependent above 1000 copies/mL.

The allowable change of VL from week to week is controlled by three model parameter settings: step size, variance and compression factor. Figure 4 demonstrates these and the different potential viral load trajectories. VL step size is considered a fixed VL parameter currently set at 0.5 $\log_{10}$ copies/mL. VL variance is magnitude dependent when VL >= 1000 copies/mL (eg an individual with a current $\log_{10}$ VL = 6 can see a larger increase or decrease in VL compared to an individual with a current $\log_{10}$ VL = 3.5). When VL < 1000 copies/mL the noise is no longer magnitude dependent, but set at 10% of current VL.

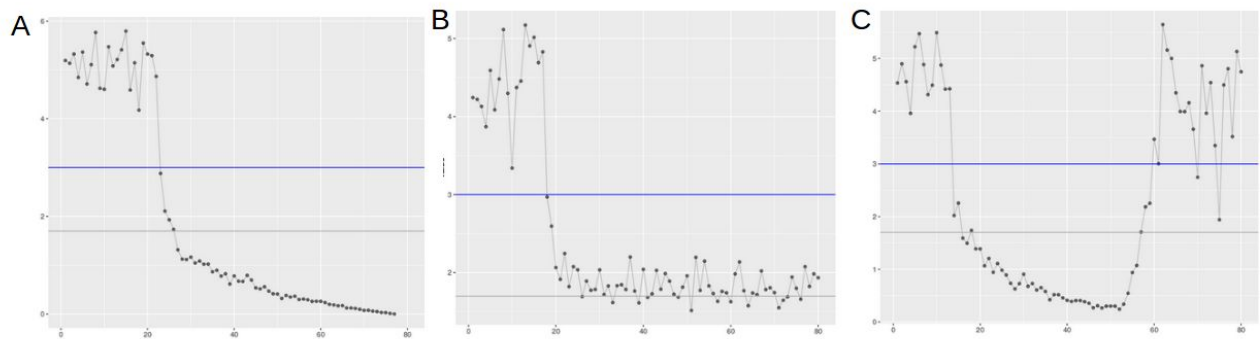

**Figure 4:** Sample viral load trajectories. Panel A: initiation of ART to sustained viral suppression. Panel B: Initiation of ART followed by sustained low level viremia. Panel C: Initiation of ART followed by sustained loss of viral control.

**MTCT risk modelling** MTCT risk is established by first setting a model input parameter that determines the per compartment overall transmission risk in an untreated population. The weekly MTCT risk is then modified by weekly maternal VL into a weekly transmission probability. Each week, a random binary value is drawn with the individual weekly transmission probability to determine if transmission has occurred. Once transmission has occurred, the individual is removed from the risk set. The model allows for three levels of VL based modification, where lower maternal VL reduced transmission risk in a stepped manner. The threshold and degree of reduction are set by internal parameters that are subject to calibration. If BF is halted, the transmission risk is set to zero. These compression factors for transmission risk are modifiable and subject to sensitivity analyses. The influence diagram for transmission risk is in Figure 5.

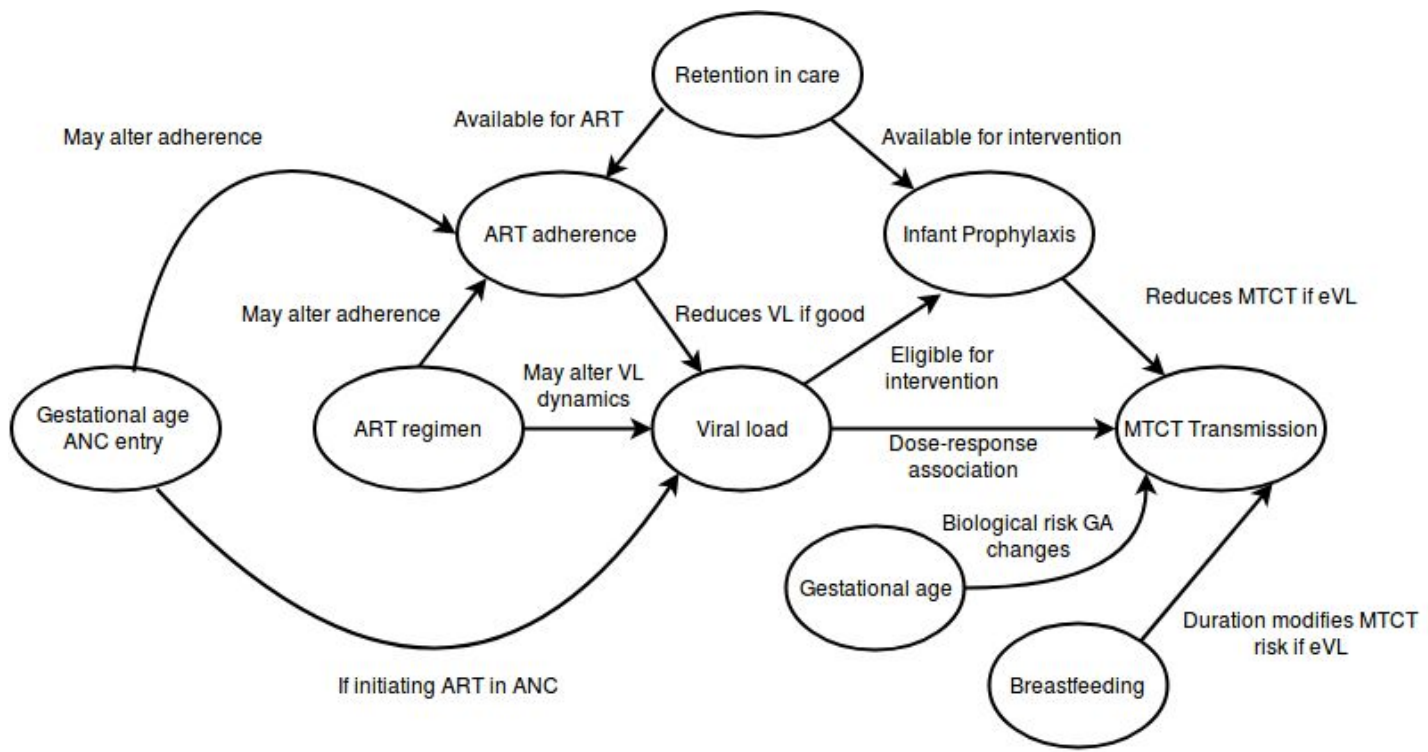

Influence diagram for primary model (individual layer): this lays out the direction of influence and multi-component influence modeled in the simulation. Eg retention in care changes availability for ART, which modifies VL, which modified MTCT transmission risk.

**Figure 5:** Influence diagram for primary model of MTCT risk

**VL monitoring** Monitoring of VL is defined as a health services level attribute, and the observed (or monitored) VL are flagged and tracked for later analysis. The basic set of assumptions are as follows: a VL can only be observed if the individual is retained in care (not LTFU) and if it is monitored, in which case the individual can have a responsive intervention applied (if the monitored VL meets a threshold). The delay between monitoring the VL and observing the VL is called the "VL cascade lag" and is an explicit parameter in the model. As an example, point-of-care VL monitoring would have a VL cascade lag equal to zero (so a raised VL could have an intervention applied in the same time step), while a health system with a VL cascade lag of eight weeks could only apply the intervention eight weeks after the VL was originally monitored. This has obvious implications in the presence of effective interventions. Individuals who have not yet entered care, or who have dropped out of care, by definition, cannot be monitored, nor can an intervention be applied to them (Figure 2). The VL monitoring scenarios are selected to provide useful information to stakeholders and span the range from weekly VL monitoring while a woman is in care, European,

WHO and African guideline monitoring schedules, and monitoring at all or some of routine ANC and PP visits (Table 2 and Supplement Table 1).

**VL monitoring optimization** The timing and frequency of viral load monitoring will be optimized for one, two or three VL monitoring tests during pregnancy and breastfeeding. This will be done varying the monitoring lag time for a number of simulated populations.

**Table 2:** Classes of modelled VL monitoring scenarios

| Scenario | Intervention options | Motivation |
|---|---|---|
| No VL monitoring | No intervention | Reflects what occurs given the population parameters and in the absence of monitoring/intervention. Used for all non-monitoring/intervention dependent calibration. |
| No VL monitoring | Intervention on pre-set schedule (ie routine ANC visits/other specified) | How non-response or scheduled interventions change outcomes. This is an evaluation of applying interventions as per a schedule instead of in response to (raised) VL. |
| Weekly VL monitoring | Trigger intervention if elevated VL detected | This is the "optimal" case and reflect monitoring to the most precise scale of the model. This is also a non-feasible and most expensive scenario, but will set an upper threshold on early detection. |
| Guidelines monitoring | Per guidelines (Supplement Table 1) | Compare existing guidelines. |
| Routine ANC visits | Trigger intervention if elevated VL detected | Evaluate care based on routine visits only. This will be done on a country specific basis as with guidelines based monitoring. |

**Intervention effect** This simulation model is mostly concerned with interventions that are applied in response to an observed raised VL, as guidelines in most SSA countries currently propose. The option is available to apply interventions across the simulated eligible population without any VL monitoring, and this scenario is modelled. Specific interventions are not modeled explicitly, but the impact of an intervention is modelled by changing an individual's probability of adherence and/or retention, which in turn impacts the current VL. Three classes of interventions are modelled: those that impact the probability of retention in care, those that directly impact the probability of adherence, and those that have a direct impact on infant transmission risk. Interventions are parameterized by the response proportion (percent of individuals that have a response to the intervention), the duration of response (which can attenuate over time), as well as any potential delay in application or effectiveness of intervention.

Maternal interventions will be applied as a theoretical intervention that can impact retention in care (reducing the weekly probability of LTFU), adherence (increasing the weekly probability of adherence) or both together. The effectiveness and duration of an intervention will be set by input parameters (eg. 50% of women will adopt intervention, and it will be effective for 7 weeks). The change to probability of retention or adherence is subject to calibration analyses and will be estimated from the literature. As interventions of this type are poorly understood, most model output looking at or using interventions will estimate effects for a range of scenarios. If interventions are responsive to raised VL, then interventions can be applied either at scheduled visits (either routine or VL monitoring visits) depending on when the VL result is made available. Alternately, interventions can be applied routinely. Maternal interventions are available to be applied in both antenatal and breastfeeding periods.

Infant interventions are available post-delivery and take only two forms, which is the possibility to apply infant prophylaxis or alter BF recommendations. Three levels of infant prophylaxis are possible: none, standard (based on 6 weeks of NVP), and enhanced (based on 12 weeks of dual drug therapy). BF can be stopped (with a lag).

**Retention in care** Retention in care is only modelled from the point of ANC entry. Women can enter continuing ART, or enter and initiate ART. HIV testing and lag between entry into ANC and initiation of ART are not modelled, assuming, as is the case in many SSA countries, that Option B+ is in place (same day ART initiation). Women are subject to a weekly probability of LTFU which is calculated using model input parameters for the overall rate of LTFU during pregnancy and the overall rate of LTFU during breastfeeding. If LTFU, a woman is assumed to be non-adherent and VL and transmission risk will rise. There is no current option for women to re-engage with care at a later time point. If a woman is not retained in care, she is not available for VL monitoring or any other health services interventions.

**Monitoring and feedback lag** In most health systems, the "turnaround time" from taking the VL sample to being able to act on the information in the sample may vary. While point of care (POC) VL is available, it is not widely distributed in SSA. Studies investigating turnaround time for routine VL monitoring in adult populations in SSA show that this can differ from a low of 72 hours in South Africa, to a high of 8-12 weeks reported in Malawi [15]. This is a modifiable input parameter. The system lag incorporates both the time to get the VL result from the lab, as well as time to re-book a woman for a visit. An example makes this clear. If a woman is scheduled to have her VL monitored at 34 weeks gestation, but the system lag is 4 weeks, the first point in time an intervention could be applied is 38 weeks (noting that this may or may not be prior to delivery). This system lag clearly impacts the utility of VL monitoring, as long system lags are likely to result in higher rates of misclassification and missed opportunities.

**Simulation Algorithm**
A brief version of the simulation algorithm is provided here.

1. Population level parameters are initialised (Population size, monitoring schedule, ANC entry characteristics, BF characteristics, baseline adherence distributions, pre-ART VL, proportion continuing ART).

2. Tracking matrices and helper functions are initialised, and individual starting parameters for the whole population are estimated from sampling distributions (eg, individual VL values at simulation start are estimated, individual time of ANC entry and delivery are set).

3. Individuals are simulated sequentially; each individual simulation is independent allowing easy parallelisation.
    a. *Conception to entry to ANC* involves maintenance of initial parameters around a mean. No VL monitoring or interventions are able to occur during this period since women are not engaged in health care. MTCT is possible and tracked.

    b. *ANC to delivery*. For women continuing ART, LTFU, VL monitoring and other interventions are able to occur. For women initiating ART, a VL suppression trajectory is started, subject to adherence, and/or LTFU. VL monitoring and interventions are able to occur. Delivery occurs at the modelled gestational age. MTCT can occur and is tracked.

    c. *Delivery until end of BF*. Adherence and LTFU rates subject to change on account of delivery. Women can breastfeed for a variable period of time. LTFU, VL monitoring and maternal interventions are available. Infant interventions are available (infant prophylaxis). MTCT can occur and is tracked.

    d. *End BF until 2 years P*P if not reached. Primarily model bookkeeping. MTCT cannot occur during this period, nor can any interventions be applied. LTFU no longer modelled (all model outcomes are estimated at end of BF at latest).

4. After full population has completed, summaries are tabulated per the model outcome measures.

**Outcome measures** The main outcome measures of interest include: the proportion of women eligible for VL monitoring, the proportion of women viremic over time, the number of monitoring VL measurements taken, the median (IQR) of viral load at time of measurement, the cumulative viremia (cVL) experienced before detection, the number of interventions applied and the total MTCT rate. These are detailed in Table 3.

**Table 3:** Model outcome measures

| **Primary outcome measure** | **Notes / time periods / groups** |
|---|---|
| MTCT per 100,000 live births | Cumulative until birth, cumulative until 6 weeks PP, cumulative until end BF |
| Proportion women with VL>1000 copies/mL among women retained in care | Before delivery; after initial viral suppression; cumulative to end BF |
| **Secondary outcome measures** | |
| Proportion of women eligible for VL monitoring | Must be retained in care at monitoring time point, and either not delivered or still BF depending on period |
| Number of VL tests done | Antenatal, postnatal, cumulative |
| Median number of VL tests done per woman | Antenatal, postnatal, cumulative |
| Proportion of women with any VL test | Antenatal, postnatal, cumulative |
| Median weeks to first VL test | Separately among continuing ART and initiating ART groups |
| Median weeks to detection of raised viremia | Product limit estimate of time to detection if monitored when VL elevated (those not monitored or not detected are considered censored at end of followup) |
| Proportion of women with VL monitoring carried out at the time of elevated VL>1000 copies/mL | Antenatal, postnatal, cumulative |
| Cumulative VL from 1st ANC until detection of VL>1000 copies/mL or 2y PP | Separately among continuing ART and initiating ART groups |
| Cumulative VL from first viral suppression until detection of VL>1000 copies/mL or 2y PP | Separately among continuing ART and initiating ART groups |

PP - postpartum, BF - breastfeeding, ANC - antenatal care, MTCT - mother-to-child transmission

**Study design and sample size considerations**

As the simulation model is complex, including a large number of parameters, a formal study design for validation and calibration [16], as well as sample size calculations, were carried out.

Two sample size calculations were considered. The first is the minimum population size for a single run of the simulation. This sample size was estimated based on the desire to estimate the primary and secondary model outcomes with specified relative precision. The MTCT transmission rates can be expected to vary from <1% in a population with good access to ART and other health services, to upwards of 40% in an untreated population. To estimate a difference of 0.005 when MTCT transmission rate is low (ie between 1% and 4%) with 95% power and an alpha of 0.008 (crudely adjusted for the six specified primary outcomes) requires a minimum sample size of N = 9157. To estimate a difference of 0.005 when MTCT transmission rate is high (eg between 30% and 40%) with 95% power and an alpha of 0.008 required a sample size of N = 177,639. This sample size was deemed too large for feasibility reasons, and precision with higher rates of MTCT transmission is of less importance, so a 1% precision was evaluated, requiring a minimum sample size of N = 44,493, rounded up to N = 50,000. A sample size of N = 50,000 will be used for all simulation runs where MTCT transmission will be estimated. Where MTCT transmission rates are not estimated or utilised (in some planned analyses investigating time to detection of raised viremia), the sample size was set at N = 10,000.

The second calculation is the number of replicates to be run for each parameter set. For this we followed the approach in Law and Kelton [17] where pilot simulations are run (varying only the random seed) until the precision of the output estimate (mean and standard deviation) is within a specified range. This procedure and results comprise part of the validation and calibration analyses and will be detailed in a forthcoming calibration document.

Other design considerations can be found in the sections pertaining to validation and calibration.

## 3 MODEL VALIDATION AND CALIBRATION

Model validation, sensitivity analyses and calibration are planned and extensive enough that full specification and results will be detailed in a different document. The main input parameters setting the population characteristics are: mean gestational age (weeks) of women at entry into ANC, the mean duration (weeks) of BF, the proportion of women entering ANC continuing ART, the total proportion of women LTFU in ANC and postpartum, the proportion of women with poor and/or mixed ART adherence.

**Internal model calibration**  Internal model validation will use a contributed data source 'MCH-ART' [18] to set the 'baseline parameters'. The model parameters (Table 4) will be altered until there is good agreement between the proportion of women viremic at different points during pregnancy and breastfeeding in the empirical data and the model generated outputs, the distribution of VL values among those not virally suppressed over time and the MTCT rate at delivery, 6 weeks postpartum and until the end of breastfeeding.

**Sensitivity analysis**  After internal model validation, sensitivity analysis will be undertaken for key internal parameters in order to better understand the scope of model variability. The baseline values will be set after internal model calibration.

The analysis plan for this section will involve calculating the parameter space of all possible combinations from parameters in Table 4, taking a subsample based on a latin square design [19] and running simulations with those settings. The MTCT and VL outcomes will be plotted for visual evaluation, and a goodness of fit measure will be calculated.

**Calibration**  The simulation model will be calibrated on key variables across a number of contributed data sources. These represent the most comprehensive and best available empirical data globally on VL in PPW. This includes datasets from South Africa (MCH-ART [18] and National Health Laboratory Services), and Botswana (the Mma-Bana study [20]), as well as the largest PMTCT trial examining ART in pregnancy set in Uganda, the PROMISE study [21,22] .

MCH-ART

A randomised control trial conducted in South Africa evaluated two delivery alternatives of HIV services to HIV infected breastfeeding women. The study was conducted in a peri-urban area in South Africa between 2015 – 2018. The current standard of care for HIV infected postpartum women, referral from the antenatal clinic at 4-8 weeks postpartum to the general adult ART services, was compared to continued maternal-focused ART services at the antenatal clinic throughout the period of breastfeeding. Maternal HIV viral suppression and retention in care up to 12 months postpartum were evaluated [18].

NHLS

Individuals undergoing HIV treatment in public sector facilities in the Western Cape province, South Africa, have routine blood samples processed by the National Health Laboratory Services (NHLS) in one of two laboratories.  NHLS data has previously been used to evaluate effectiveness of government funded HIV programs.

Mma-bana

A randomised control trial conducted in Botswana from July 2006 to May 2008, followed mother infant pairs on three different ART regimens. HIV infected ART naive women (n = 730) between 26 and 34 gestation weeks were randomised to either abacavir, zidovudine, lamivudine (Arm A) or lopinavir–ritonavir, zidovudine–lamivudine (Arm B) and followed up to 6 months postpartum. The third arm was observational, where women with a CD4 cell count below 200 received nevirapine–zidovudine–lamivudine [20]. All women received multiple VL tests during follow up.

PROMISE

PROMISE (1077BF/FF) was a randomised control trial comparing the risk of placental malaria among lopinavir and efavirenz study arms. The study was conducted in rural Uganda, from December 2009 to March 2013 and enrolled 389 HIV infected pregnant women [21]. A subset (n=200) of the women that participated in the PROMISE P2 study, were invited to partake in an observational study (BC2). The study assessed maternal viral suppression and retention in care up to 5 years postpartum [22].

Each simulation run to calibrate VL over time will be evaluated on each data set using a goodness of fit test described below.

The basic algorithm is:
1. Identify large parameter set to start (selected subset from sensitivity analysis).
2. Run simulations for all parameter sets in 1.
3. Select parameter sets to carry forward IF % viral suppression at delivery is within 5% of the data source measure of % viral suppression at delivery.
4. If NO postnatal data then compare % viral suppression at antenatal time point between entry to ANC care and delivery. This will vary depending on the data set, and retain parameter sets within 5% of data source.

5. If postnatal data is available, then compare % viral suppression at 3m postpartum and retain parameter sets within 5% of data source.

The retained parameter sets will be evaluated with a goodness of fit test to find the 'best' parameter set for a given data source. The root mean square error (RMSE) for the goodness of fit test is found in equation (1).

$$\text{RMSE} = \sqrt{\sum_{t=1}^{T} \frac{(VS_P(t) - VS_D(t))^2}{T}} \qquad (1)$$

The RMSE is calculated as the square root of the average of the squared difference between observed and projected proportion of women with VL <1000 copies/mL at each time point available in the source data, denoted by VS(t).

**4 CONCLUSION**

The detailed rationale, motivation and design considerations for the VL-SiM PPW simulation model are presented. Statistical considerations around design, sample size and feasibility are addressed and the current model algorithm is presented.

To our knowledge, there are no individual stochastic simulation models that incorporate longitudinal adherence and VL trajectories in pregnant and breastfeeding HIV+ women. This model presents a novel opportunity to simulate patterns of VL in an important population to understand the role and impact of VL monitoring on detection and intervention for the prevention of vertical transmission.

**Table 4:** Model input and other parameters for sensitivity and calibration analyses

| Parameter | Description |
| --- | --- |
| Mean gestational age at ANC entry | Impacts the average duration of treatment for those initiating ART during pregnancy |
| Percentage women initiating ART | Modifies the population proportion continuing or initiating ART during |
| during antenatal care | pregnancy |
| Proportion 'full' adherent | Controls the percentage of women considered to be adherent sufficient to sustain viral suppression |
| Proportion 'partial' adherent | Controls the percentage of women with adherence that is not always sufficient to sustain viral suppression, but who have intermittent patterns of adherence |
| Mean duration breastfeeding | Modifies the average duration of breastfeeding |
| Percentage women LTFU | Controls the overall proportion of women LTFU (antenatal and postnatal) |
| Distribution 'pre-ART' viral load | Alters the incoming VL distribution of women initiating ART |
| Viral load 'step size' | Controls the mean allowable change in VL from week to week |
| Viral load 'noise' | Controls the magnitude of additive noise on change in VL from week to week |
| Adherence 'noise' | Controls the magnitude of additive noise on change of adherence from week to week |
| MTCT risk period 1 | Early antenatal vertical transmission risk in an untreated population |
| MTCT risk period 2 | Peripartum vertical transmission risk in an untreated population |
| MTCT risk period 3 | Early breastfeeding vertical transmission risk in an untreated population |
| MTCT risk period 4 | Late breastfeeding vertical transmission risk in an untreated population |
| MTCT risk compression factors | Modifies the VL - risk dose dependence |

**Table 5 :** Details of guidelines based monitoring at the time of writing. Additional monitoring guidelines may be added as they become available.

| Guideline (year) [ref] | VL monitoring time points antenatal | VL monitoring time points breastfeeding |
|---|---|---|
| WHO (2016) [1] | If initiating ART: 6m, 12m, then annually and at 34-36 weeks GA | Annually |
| | If continuing ART: annually from ART initiation date and at 34-36 weeks GA | |
| USA DHHS (2018) [23] | If initiating ART: 2-4 weeks after ART initiation, then monthly until VS, then 3 monthly (if VL,50 copies/ml) plus at 34-36 weeks GA<br><br>If continuing ART: 1st ANC visit and routinely every month, move to monitoring every 3 months (if VL<50 copies/ml) plus 34-36 weeks GA | BF not recommended in this population |
| UK (2018) [24] | If initiating ART: 2-4 weeks post initiation, at least every trimester and at 36 weeks GA<br><br>If continuing ART: at least every trimester and at 36 weeks GA | BF not recommended in this population |
| South Africa (2015) [4] | If initiating ART: 3m, 6m then every 6m<br><br>If continuing ART: first ANC, every 6m | Continue every 6m until end of BF |
| Kenya (2016) [5] | If initiating ART: 6m, then every 6m<br><br>If continuing ART: first ANC, then every 6m | Continue every 6m until end of BF |
| Malawi (2016) [25] | If initiating ART: 6m, then every 24m<br><br>If continuing ART: every 24m | Continue every 2y until end BF |
| Uganda (2016) [26] | If initiating ART: 6m post ART, then annually<br><br>If continuing ART: first ANC, 6m, then annually | Continue annually until end BF |
| Zambia (2018) [6] | If initiating ART: 6m, then every 6m plus at 34-36 weeks GA<br><br>If continuing ART: first ANC, then every 6m plus at 34-36 weeks GA | Continue every 6 m until end BF |

BF: breastfeeding; ANC: antenatal care; ART: antiretroviral therapy; WHO: World Health Organisation; GA: gestational age

**Ethics**

This work has been carried out under approval from the University of Cape Town Faculty of Health Sciences Human Research Ethics Committee (HREC: 865/2016).

**Funding**

Research reported in this publication was partially supported by the Eunice Kennedy Shriver National Institute of Child Health & Human Development of the National Institutes of Health under award number R21HD093463. The content is solely the responsibility of the authors and does not necessarily represent the official views of the National Institutes of Health.

TG is supported by the Medical Research Council of South Africa in terms of the National Health Scholars Programme from funds provided for this purpose by the National Department of Health/Public Health Enhancement Fund and by the South African Department of Science and Technology/National Research Foundation (DST-NRF), Centre of Excellence in Epidemiological Modelling and Analysis (SACEMA), Stellenbosch University, Stellenbosch, South Africa.

**Acknowledgements**

We thank the many researchers who have shared data and ideas with us as we developed this work, and the participants in all of the research studies who contribute their time, samples and knowledge.